\documentclass{aa}
\usepackage{graphicx}
\def\approxgt{\mathrel{\hbox{\rlap{\lower.55ex \hbox {$\sim$}}
        \kern-.3em \raise.4ex \hbox{$>$}}}}
\def\approxlt{\mathrel{\hbox{\rlap{\lower.55ex \hbox {$\sim$}}
        \kern-.3em \raise.4ex \hbox{$<$}}}}
\begin{document}
   \title{A BeppoSAX view of the Centaurus Cluster}


   \author{Silvano Molendi
          \inst{1}
          \and
           Sabrina De Grandi
           \inst{2}
          \and
           Matteo Guainazzi
           \inst{3}
          }

   \offprints{S.Molendi}

   \institute{
         Istituto di Fisica Cosmica, CNR, via Bassini 15,
         I-20133 Milano, Italy
   \and
         Osservatorio Astronomico di Brera, via Bianchi 46,
        I-23807 Merate (LC), Italy
   \and
         XMM-Newton SOC, VILSPA ESA, Apartado 50727, 
        E-28080 Madrid, Spain\\
              }

   \abstract{
 We present results from the analysis of a BeppoSAX
 observation of the Centaurus Cluster.
 The radial metal abundance profile shows evidence of 
 a large enhancement in the core, where Ab $>$ 1 (sol. units). 
 The temperature map indicates that the cluster is characterized by   
 a gradient oriented in the NW/SE direction,
 with cooler gas in the NW and hotter gas in the SE 
($\Delta kT \sim 1$ keV).
 In hard X-rays , where the PDS detects emission with a flux of 
$2.2 \times 10^{-11}$ erg/cm$^2$/s in the 20-200 keV band, 
the spectrum lies above the  extrapolation of the cluster thermal 
emission. 
 We discuss several possible interpretations for the hard
 excess finding that none is completely satisfactory.
   }

   \maketitle
%

\section{Introduction}

The Centaurus Cluster is amongst the nearest (z=0.0114) and
brightest clusters in the X-ray sky. It has been looked at 
with all major X-ray missions. 
In recent years, observations with the ASCA  satellite have 
produced  a number of interesting results.
Analysis of the core of the cluster has   
provided evidence for a high metal concentration,  
the measured abundance exceeds the solar
value (Fukazawa et al. 1994), and for a cool emission component
(Fukazawa et al. 1994, Finoguenov et al. 2001, 
Furusho et al. 2001).
Churazov et al. (1999) and Furusho et al. (2001)
have detected asymmetric temperature variations 
on the few hundred kpc scales.
The fact that the hot region located SE of the cluster 
is roughly coincident with the position of NGC4709, 
which is the dominant galaxy of the Cent45 subcluster,
indicates that the cluster is probably undergoing a 
merger event.
In this paper we present results from the analysis of a
BeppoSAX (Boella et al. 1997a) observation of the 
Centaurus cluster.
We use the MECS experiment (Boella et al. 1997b), which is 
characterized by a spatial resolution a factor of $\sim2$ 
better than that of the ASCA instruments, to investigate the
temperature and metal abundance structure of Centaurus.
We also use the PDS experiment (Frontera et al. 1997), 
to study the hard X-ray emission.
The paper is organized as follows: 
in Sect.~2 we describe the data reduction. 
In Sect. ~3 we present the  spatially resolved spectroscopy
performed with the MECS instrument.
In Sect.~4 we describe the PDS spectrum.
In Sect. ~5 we discuss our results and compare them to previous
findings.
Confidence intervals are at the 68\% level for one interesting parameter, 
unless otherwise stated.
We assume H$_o=$50 km s$^{-1}$Mpc$^{-1}$.

\section{Observations and Data Preparation}
BeppoSAX observed the Centaurus Cluster between 
2000 July 3 (15:03:31 UTC) and July 5 (11:21:44 UTC). All the 
instruments were operated in standard photon-counting
direct modes, the exposure times were respectively 70 ks for MECS and
 31 ks for PDS.  
Data reduction followed standard
procedures as described, {\it e.g.}, in
Guainazzi et al. (1999). In this 
paper only data from the Medium Energy
Concentrator Spectrometer (MECS, 1.8--10.5~keV) and from the Phoswitch
detector System (PDS, 13--200~keV) will be presented.

We have analyzed the data from the MECS2 and MECS3 
separately. The SAXDAS package under 
FTOOLS environment has been applied to produce equalized and 
linearized MECS event files.  Using the information contained in 
the housekeeping files we have rejected all events which have 
occurred at times when the instantaneous pointing direction 
differed by more than $10^{\prime\prime}$ from the mean 
pointing direction. 

The PDS data have been reduced using a
variable Rise Time threshold method, which
allows an increase of the signal-to-noise
(S/N) ratio for sources fainter than 1~mCrab.
The corresponding normalization factor (0.8;
Fiore et al. 1999) has been applied in all
the spectral fits discussed in this paper.


\section{Spatially resolved spectroscopy}

Spectra have been prepared for analysis following the guidelines
described in De Grandi \& Molendi (2001) and De Grandi \& Molendi
(2002), below we summarize the various operations  which have been 
performed.  
 
Spectral distorsions introduced by the energy dependent PSF
are taken into account using the {\sc Effarea} program
publicly available within the latest {\sc Saxdas} release.

The temperature  and metal abundance radial profiles are truncated  
at the last bin where the source counts exceed 
$30\%$ and $40\%$ respectively of the total (i.e.  source$+$background) 
counts. 
The energy range used for the spectral fitting is 2-10 keV with the 
following exceptions . 
 In the outermost bins, if the source counts drop to less than 
$50\%$ with respect to the total counts, we restrict the energy range 
to 2-8 keV, to avoid possible distortions from the hard MECS 
instrumental background. 
A strongback correction has been applied to the effective area
for the $8^\prime-12^\prime$ annulus, in this bin we 
restrict our analysis to the range 3.5-10 keV to avoid the low energy
part of the spectrum where our correction is less reliable.  All other
regions of the detector covered by the strongback have been
appropriately masked and the data rejected.
 
The background subtraction has been performed using spectra extracted 
from blank sky events files in the same region of the detector as the 
source. 

\subsection{Radial Profiles}

The cluster has been divided into concentric annuli centered on the 
X-ray emission peak. Out to $8^{\prime}$ we accumulated spectra from 4 
annular regions each $2^{\prime}$ wide, beyond this radius we 
accumulated spectra from annuli $4^{\prime}$ wide.   

We have fitted each spectrum with a single temperature model (MEKAL code in
XSPEC) absorbed 
by the galactic line of sight hydrogen 
column density, N$_{\rm H}$, of $8.06\times 10^{20}$ cm$^{-2}$ derived
from HI maps ( Dickey \& Lockman 1990). 
All spectra, with the exception of the one associated to the innermost bin,
can be adequately fit with the above model. 
In the innermost region the single temperature model leaves a rather large
excess above $\sim$5 keV. 
Since ASCA (e.g. Makishima et al. 2001 ) and XMM-Newton (Molendi \& Pizzolato 2001) studies of 
the cores of centrally peaked clusters have shown that the spectra 
accumulated from
these regions are generally better fitted by multi temperature models than 
by single temperature models, we have performed a new fit with a model 
including a second MEKAL component. 
The metal abundace of this second 
component  has been linked to the metal abundance of the first. 
Since the temperature of the hotter component is not well constrained we 
have fixed it to 4 keV, which is approximatively the largest temperature 
we measure in our profile (see Fig. 1) and in rough agreement with a
maximum temperature of 4.4 keV 
found  by Ikebe et al. (1999) from the analysis of ASCA data.
The two temperature model yields a substantially better fit than the 
one temperature model, the improvement is statistically significant
at more than the 99\% level according to the F-test. 
Note however that the two temperature modeling should be regarded as a 
rough attempt to reproduce the complex spectrum resulting from the superposition 
of the cooler emission coming from the core of the cluster with the hotter 
emission coming from regions further out, which happen to be on the 
same line of sight. Given the limitations of the MECS instrument both 
in terms of spatial resolution and  of band coverage, the MECS data 
does not allow a more detailed analysis of the temperature structure in 
the core of Centaurus.
  
%
   \begin{figure}
   \centering
   \includegraphics[bb=55 68 617 824,angle=-90,width=11.5cm]{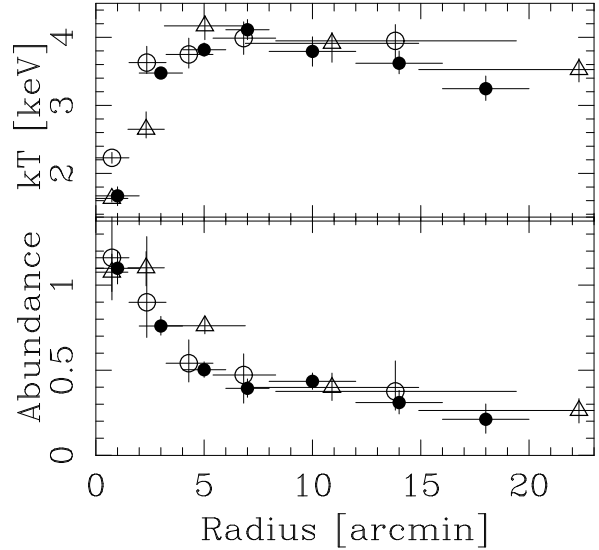}
      \caption{
{\bf Top Panel}: Projected radial temperature profile for the Centaurus
Cluster. Filled circles indicate temperatures obtained from BeppoSAX MECS 
data. Open circles and open triangles represent  the temperature 
profile derived from ASCA data respectively by White (2000) and by 
Finoguenov et al. (2001) 
{\bf Bottom Panel}: projected radial abundance profile. Symbols 
as for Top Panel.
              }
   \end{figure}
%
  
In Fig. 1  we show the temperature and abundance 
profiles obtained from the spectral fits with single temperature models.
The values reported for the innermost bin
are respectively the lower temperature  and the metal abundance obtained with  
the 2 temperature model discussed above.

The temperature profile in this, as in other centrally peaked clusters, 
is found to rise with increasing radius, roughly leveling off 
at the cooling radius , r$_{\rm c} \sim  4^\prime$ (Peres et al. 1998).
Beyond  r$_{\rm c}$ the profile shows a  drop.
When fitting data for r$>4$ arcmin, a declining line gives a slightly  
better fit than a constant (the improvement is significant only at the 
75\%  level according to the F-test). Note however that if we concentrate 
only on the last 4 data points (r$>6$ arcmin) the improvement is found to be
significant at more than the 99\% level. 

The metal abundance profile shows a striking excess in the innermost bin.
We remark that the measurement in this bin is not particularly sensitive 
to the spectral model, indeed  the single temperature and 2 temperature models
described above give very similar results. Metal abundance excesses in the core 
of so-called cooling-flow clusters have been found in virtually all BeppoSAX
observations (see De Grandi \& Molendi 2001), however the value we find for 
Centaurus is the largest  measured so far and the only one to exceed the 
solar value.
The abundace drops rapidly  with increasing radius reaching a value of 
$\sim$ 0.4 solar units at $\sim 7^\prime$ from the center. From here on
the abundance declines gently, reaching a value of $\sim$ 0.2 solar units
at $\sim 18^\prime$ from the center.
While the abundance excess in  the core is most likely associated to
enrichment from the cD galaxy NGC 4696 itself, the gentle declining 
profile observed further out could be due to enrichment by galaxies 
in the core of Centaurus similar to what has been found by Ezawa et al. 
(1997) in the case of AWM 7.

\subsection{Maps}
   \begin{figure}
   \centering
   \includegraphics[angle=0,width=8.0cm]{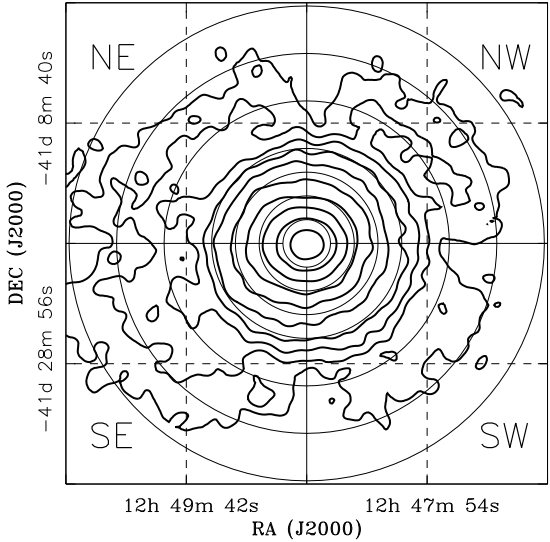}
      \caption{
BeppoSAX MECS image of the Centaurus Cluster. Contour levels are indicated
by the thick lines. The thin lines show how the cluster
has been divided to obtain temperature and abundance maps.
Note that no correction for the strongback has been 
applied to the image.}
   \end{figure}
%

As shown in Fig. 2, we have divided the MECS image of the Centaurus 
cluster into 4 sectors: NW, SW, SE and NE, each sector has been 
divided into 6 annuli with bounding radii,
2$^{\prime}$-4$^{\prime}$, 4$^{\prime}$-6$^{\prime}$, 
6$^{\prime}$-8$^{\prime}$, 8$^{\prime}$-12$^{\prime}$, 12$^{\prime}$-16$^{\prime}$ 
and 16$^{\prime}$-20$^{\prime}$. 
In Fig. 3 and 4 we show respectively  the temperature and abundance 
profiles obtained from the spectral fits for each of the 4 sectors.
Note that the abundance profiles extend only to 16$^{\prime}$ because
some of the spectra in the  16$^{\prime}$-20$^{\prime}$ 
annulus do not meet the more stringent conditions required  
to accept an abundance measurement. 
To ease the comparison between measurements in different sectors we have 
drawn a constant line for a temperature of 4 keV  and a metal abundance 
of 0.5 in Fig. 3 and 4 respectively.

\begin{table*}
\begin{center}
\begin{tabular}{lcrccrc}
\hline
Sector &    kT       & $\chi^2$/dof & Prob.$^a$         & Abundance   & $\chi^2$/dof & Prob.$^a$       \cr
       &  (keV)      &          &                   &(Solar Units)&          &                        \cr
\hline
\hline
All    & 3.6$\pm$0.1 &   23.2/5   &$3.1\times 10^{-4}$& 0.47$\pm$0.02&   42.6/5  &$4.5\times 10^{-8}$ \cr  
\hline
NW     & 3.2$\pm$0.1 &   10.0/5   &$7.5\times 10^{-2}$& 0.40$\pm$0.05&   25.6/4  &$3.9\times 10^{-5}$ \cr
SW     & 3.8$\pm$0.1 &   13.0/5   &$2.4\times 10^{-2}$& 0.47$\pm$0.05&    6.5/4  &$1.7\times 10^{-1}$ \cr
SE     & 4.0$\pm$0.1 &    5.0/5   &$4.2\times 10^{-1}$& 0.46$\pm$0.04&    6.8/4  &$1.4\times 10^{-1}$ \cr
NE     & 3.6$\pm$0.1 &    5.7/5   &$3.4\times 10^{-1}$& 0.57$\pm$0.05&    6.0/4  &$2.3\times 10^{-1}$ \cr
\hline\hline
\end{tabular}
\end{center}

\noindent
$^a$Probability for  the observed distribution to be drawn from a constant parent distribution.
\caption{
Best fit constant values for the temperature and
the abundance for the azimuthally averaged radial
profiles (All) and for the profiles of the 4 sectors (NW, SW, SE and NE)}
\label{tab1}
\end{table*}

In Table 1 we report the best fitting constant temperatures and 
abundances for the profiles shown in Fig. 3 and 4.

   \begin{figure}
   \centering
   \includegraphics[angle=-90,width=8.0cm]{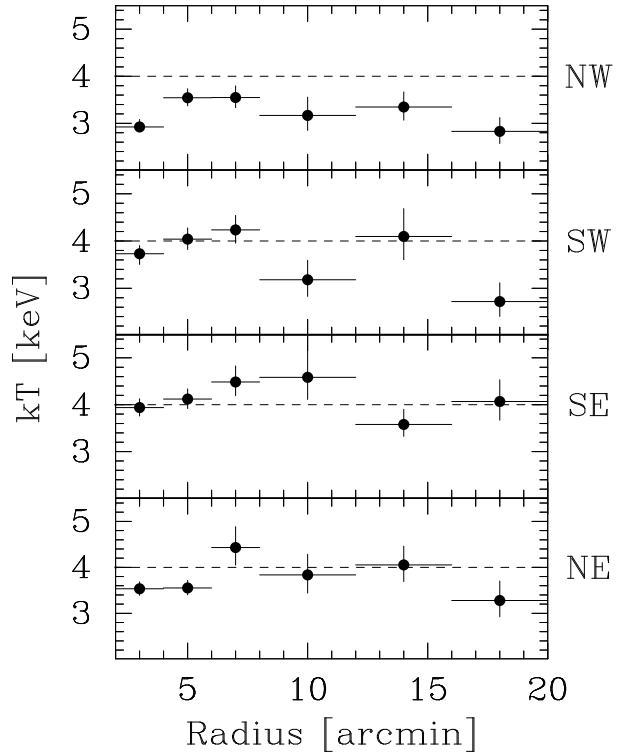}
      \caption{
Radial temperature profiles for the NW sector (first panel), the SW sector
(second panel), the SE sector (third panel) and the NE sector (forth
panel).
The dashed horizontal lines at 4 keV are drawn to ease the comparison
between different sectors.  
              }
   \end{figure}
%

The mean temperature of the NW sector is the smallest, $3.2\pm0.1$ keV, 
while that of the SE is the highest, $4.0\pm0.1$ keV. The difference in 
these values is significant at more than the $5\sigma$ level indicating   
that the cluster is characterized by an azimuthal temperature gradient.
There appears to be a cooler region NW  and a hotter region SE of the 
center.
A fit to the temperatures of the  4 sectors in the 
annulus with bounding radii 2$^\prime$-4$^\prime$ with a constant, yields
$\chi^2=20.1$ for 3 d.o.f., with an associated probability for
the temperature to be constant of $1.6\times 10^{-4}$, indicating that
the azimuthal temperature gradient extends to the smallest scales accessible 
to the MECS. The NW and the SE sectors are again respectively the coolest 
and the hottest ones.
   \begin{figure}
   \centering
   \includegraphics[angle=-90,width=8.0cm]{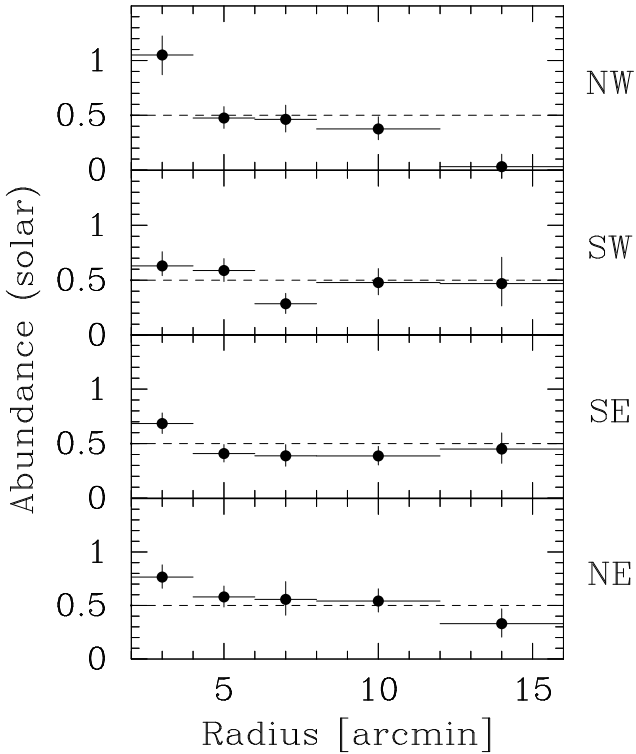}
      \caption{
Radial abundance profiles for the NW sector (first panel), the SW sector
(second panel), the SE sector (third panel) and the NE sector (forth
panel).
The dashed horizontal lines at 0.5 solar units are drawn to ease the 
comparison between different sectors.  
              }
   \end{figure}
%

Our data does not show compelling evidence for azimuthal abundance 
gradients. There is however a hint of an abundance excess in the 
NW sector of the 2$^\prime$-4$^\prime$ annulus, which has an abundance 
larger than those of all other sectors of the same annulus and 
comparable to that of the 0$^\prime$-2$^\prime$ annulus.
Interestingly this is also the sector showing the smallest  
temperature in the 2$^\prime$-4$^\prime$ annulus.

\section{The hard X-ray spectrum}

In the (1.3$^{\circ}$)$^2$
PDS field-of-view a signal at about 3.6$\sigma$ level
was detected.
In the 13--200~keV  energy band the net background-subtracted
count rate is $0.161 \pm 0.044$. The error
includes the typical
systematic uncertainties on the PDS background
subtraction algorithm (see the discussion in Matt et al. 1997). A
simple power-law fit yields a $\Gamma = 1.5 \pm^{1.4}_{0.8}$
($\chi^2 = 15.1/22$~dof).
A fit with a bremsstrahlung model is equally good,
but only a 90\% lower limit of 30~keV on the
electron temperature can be set.
The observed 20-200~keV flux is
$2.2 \times 10^{-11}$~erg~cm$^{-2}$~s$^{-1}$,
corresponding to a rest-frame luminosity of
$9.4 \times 10^{42}$~erg~s$^{-1}$ at the redshift
of NGC~4696.

The extrapolated flux
in the 2--10~keV band is $4.2 \times 10^{-12}$~erg~cm$^{-2}$~s$^{-1}$. 
The probability that the PDS detection is due to a serendipitous 
source in the PDS field of view is $\sim 5\%$, according to the 
Cagnoni et al. (1998) LogN-LogS. This is actually a lower limit, given 
the enhanced density of AGN expected in the cluster environment. A more 
accurate estimation, which takes into account the LogN-LogS of the 
Centaurus Cluster, is discussed in Sect.5
The PDS data points lie well above the extrapolation
of the MECS thermal best-fits
for all the annuli.
The PDS emission could be due to diffuse emission 
related to the merger  in Centaurus, however the absence 
of a radio halo argues against this possibility.
Alternatively the PDS excess could come
from a hard point-like source, which is either  unresolved
by the MECS, or lies beyond the MECS field of view (we recall that the
MECS has a circular FOV with a radius of $\sim$ 22 arcmin).
In order to investigate the  point source alternative, we have retrieved
from the public archive the data of a
{\it Chandra} observation of NGC~4696,
performed on May 22 2000.
The ACIS-S was operated
in the standard configuration, being exposed for
a total exposure time of 32.1~ks. The
{\it Chandra} observation covers, however,
only a small part of the PDS field-of-view, the ACIS-S FOV is a 
square 8 arcmin on the side.
We have
extracted an image of the whole {\it Chandra}
field from the
"level 2" event list. We employed the energy
band between 5 and 10~keV to minimize the contribution
of the cluster emission and therefore optimize
the detection likelihood for point sources. The
brightest detected point source has coordinates:
$\alpha_{2000}$=12$^h$49$^m$06$^s$.2,
$\delta_{2000}$=-41$^{\circ}$17$\arcmin$52$\arcsec$, and
net count rate of $(6.8 \pm 1.9) \times 10^{-4}$~s$^{-1}$.
The extrapolation of this
rate into the PDS energy bandpass,
assuming a $\Gamma = 1.5$ power-law, is
$4 \times 10^{-3}$~s$^{-1}$
therefore about 40 times lower
than the PDS count rate measured
during the BeppoSAX observation. The {\it Chandra}
count rate
3-$\sigma$ upper limit at the position of the optical nucleus
of NGC~4696 is $9 \times 10^{-5}$~s$^{-1}$.
If the detection in the PDS is due to the
emission of a strongly absorbed active
nucleus at the position of the optical nucleus, 
the column density, $N_{\rm H,nuc}$ must be $\simeq 3.3 \times
 10^{24}$~cm$^{-2}$.

\section{Discussion}
In Fig. 1 we have compared our radial temperature and abundance 
profiles with temperature and abundance  profiles derived
from ASCA data independently  by White (2000) and
Finoguenov et al. (2001). Overall the BeppoSAX temperature profile
is in good agreement with the ASCA profiles. There are some differences
within the innermost 4 arcminutes  amongs the three profiles. 
Such differences are likely related to  the different spectral bandpass
over which the spectral fitting of the data has been performed. 
At radii larger than 4$^\prime$ the three different temperature
measurements yield results which are comfortably consistent.  
The abundace profiles are also in good agreement. There is one apparent
exception, i.e. the abundace measurement at 5$^\prime$  by Figuenov et.
al (2001). The difference here however is due to the fact that the bin 
adopted by Finoguenov et.al (2001) is substantially larger than those
used in White (2000) and in our own analysis. Thus the spectrum 
fitted by Finoguenov et. al (2001) also includes emission from further
in where the abundance is larger.

Our temperature map shows evidence of a gradient in the NW/SE direction,
the cooler gas is in the NW sector and the hotter gas in the SE. 
Interestingly the ASCA temperature maps of Centaurus reported 
by Churazov et al. (1999) and Furusho et al. (2001) show evidence  
of a temperature gradient oriented as ours.   

We present here an abundance map for Centaurus. It does not provide 
compelling evidence for azimuthal abundance  gradients. The only 
result of some interest is the hint of an excess in the NW sector
of the 2$^{\prime}$-4$^{\prime}$ annulus. Interestingly this  
sector is also characterized by the smallest  temperature  in the   
2$^{\prime}$-4$^{\prime}$. Thus both in terms of temperature and
of metal abundance the emission in this sector is more similar
to the emission from the innermost region than to that of the 
other sectors of the same annulus. It is tempting to speculate 
that gas in this sector may be more closely related to gas 
in the innermost region than to gas at similar radial distances in other
sectors.

Shortly after this paper was submitted, Sanders et al. (2002) published 
a detailed analysis of the Chandra observation of the Centaurus core. 
Thanks to its better spatial resolution Chandra can investigate temperature
and abundance variations on scales smaller than those accesible to the
BeppoSAX MECS. It is therefore quite reasuring that the presence a of cooler 
and metal richer region extending out to $\sim$ 3 arcmin NW of the core is 
confirmed by Chandra. Indeed it proves that the MECS are capable of 
detecting variations in temperature and abundance down to scales of 
a few arcmin.

Apparently unrelated to the thermal emission imaged by the MECS, a signal at
the 3.6$\sigma$ level is measured in the 13-200~keV energy band with the PDS
instrument. It is impossible from the PDS data alone to establish the origin of
this emission. 
A possible explanation could be that we are observing diffuse hard X-ray
emission associated to the cluster merger,
however, the fact that no radio halo has  been detected in the Centaurus 
cluster, argues against this possibility. 

The X-ray luminosity associated to the PDS detection exceeds
$10^{43}$~erg~s$^{-1}$. This would favor emission from a strongly absorbed
active nucleus. If this hypothesis is true, the comparison with a {\it non
simultaneous} {\it Chandra} observation constraint the nuclear absorbing column
density to be $\simeq 3.3 \times 10^{24}$~cm$^{-2}$.
This hypothesis is, however, not supported by evidences in other wavelength. No
highly ionized emission lines has ever been discovered in optical spectroscopy
of the NGC~4696 nucleus. Although this is not a conclusive proof (other cases
of optically "dull" AGN are known and being discovered, the prototype of this
class being NGC~4945; Matt et al. 2000; see also the discussion in Mushotzky 
et al. 2000), there is no wavelength in the NGC~4696 spectral energy 
distribution where the reprocessed X-rays from the absorbed active nucleus 
seem to emerge. The total far infrared (40--120~$\mu$m) luminosity is $\sim 1.2
\times 10^{42}$~erg~s$^{-1}$ only (de Jong et al. 1990). The IR colors as
measured by IRAS are the coolest among the elliptical galaxies of the 
Jura et al. (1987) sample. The possibility of a serendipitous source in the 
PDS cannot be excluded a priori, given the lack of simultaneous imaging 
coverage of its field of view. 
We have tried to estimate the likelihood of such an event using the
ROSAT All Sky Survey (RASS). The extrapolation of the PDS spectrum in
the ROSAT PSPC band yields a count rate of 0.12 cts s$^{-1}$.
Within the PDS field of view, no RASS source is detected at this
intensity level. However, X-ray absorbed sources could be easily missed 
in the ROSAT Centaurus Cluster census. One can estimate the expected 
density of {\it all} X-ray sources in Centaurus by looking at a wider 
area to increase the counting statistics, and correcting the results 
for the fraction of missed absorbed objects.
In a 3$^{\circ}$ radius circular area, 6 sources
brighter than the above PSPC count rate limit are detected, corresponding to
about 1.1 source per (1.3$^{\circ}$)$^2$ circle, even if no
correction for the obscured objects is done.
Of course, the extrapolation of the X-ray counts along the almost two
order of magnitudes in energy which separate the
energy bandpasses of the BeppoSAX/PDS and the ROSAT/PSPC
introduces rather large uncertainties in this estimate.
Alternatively, one may employ the LogN-LogS of the Centaurus Cluster
(Jerjen \& Dressler 1997) to estimate the number of potential AGN
fields.
The PDS measurement corresponds to a 1\,keV flux density of 500\,mJy.
Assuming a typical quasar optical-to-X-ray luminosity
ratio of 0.5 dex (Elvis et al. 1994),
this flux density corresponds to a B magnitude of 10.6.
The number of galaxies with $m_B\approxgt 10.6$ in the 2.2 square
degrees survey of Jerjen \& Dressler (1997) is $\simeq 12$. 
Even assuming a 10\% fraction of active galaxies, this implies 
that the number of expected AGN in the PDS field of view is $\simeq 0.4$.
Both the above tests suggest that the PDS detection
may be due to a serendipitous X-ray hard field AGN.

\begin{acknowledgements}
The BeppoSAX satellite is a joint Italian-Dutch program.
\end{acknowledgements}

\end{document}